\documentstyle[11pt,paspconf,epsf]{article}

\begin{document}

  \title{Determination of the LMC distance modulus from 
         the Classical Cepheid Period-Luminosity relation}

  \author{Xavier Luri, J. Torra, F. Figueras}
  \affil{Departament d'Astronomia i Meteorologia, 
         Universitat de Barcelona, Avda. Diagonal
         647, E-08028, Barcelona, Spain}

  \author{A.E. G\'omez, M.J. Goupil \& J.P. Beaulieu}
  \affil{Observatoire de Paris-Meudon
         D.A.S.G.A.L., URA CNRS 335, 
         F92195  Meudon CEDEX, France}


\begin{abstract}
  The distance modulus of the Large Magellanic Cloud (LMC) is the first
  step on the determination of the cosmic distance scale. In this paper
  it is obtained from a new calibration of the classical Cepheid 
  Period-Luminosity (PL) relation. 

  Our new calibration is obtained by applying the LM method -- a parametric 
  method designed to obtain luminosity calibrations from astrometric 
  information -- to the Hipparcos data for Classical Cepheids. The 
  resulting distance modulus is $18.35^{m}\pm 0.13$.
\end{abstract}


\section{Introduction}

The purpouse of this paper is twofold. On the one hand, it is intended to
be a complement of the Arenou \& Luri paper in this volume, presenting an
example of application of a parametric method for luminosity calibrations. 
On the other hand, a new result for the Cepheid PL relation using
Hipparcos data is presented, and the LMC distance modulus is deduced from
it.

\section{The sample} \label{sec_sample}

Our working sample was formed by selecting all classical cepheids contained
in the Hipparcos catalogue (ESA, SP-1200), even first overtone pulsators.
The resulting sample contains 238 stars. 
  
All data (including periods) were taken from the Hipparcos catalogue,
except the radial velocities, that were taken from the \emph{Hipparcos
Input Catalogue} (Turon et al. 1992).

\section{The LM method} 

As described in Arenou \& Luri (this volume) parametrical methods can be
applied to obtain unbiased luminosity calibrations. In our case, to obtain
the PL relation of classical Cepheids we have used the LM method. This
method, using Maximum-Likelihood estimation and specifically designed for
luminosity calibrations, is fully described in Luri et al. (1996). Its main
advantages are:

\begin{itemize}

  \item All the available information for each star can be used:
        apparent magnitude, position, trigonometric parallax, proper motions 
        and radial velocities as well as other relevant astrophysical 
        parameters (color index, period, etc.).

  \item It has been designed to fully exploit Hipparcos data for
        luminosity calibrations.

  \begin{itemize}

  \item The observational selection criteria used to construct the sample
        of stars are taken into account. This is essential to obtain
        unbiased results (Brown et al., 1997 Arenou \& Luri, this
        volume).

  \item The effects of observational errors are also taken into account, 
        thus avoiding biases (Arenou \& Luri, this volume).

  \end{itemize}

  \item It gives at the same time the kinematical properties and spatial 
        distribution of the sample.

  \item It is able to handle inhomogeneous samples composed of
        groups of stars with different luminosity, kinematical 
        or spatial characteristics. 

  \item It provides an individual distance estimate with optimal 
        statistical properties (Arenou \& Luri, this volume).

\end{itemize}

\section{Modeling of the sample}

As described in Arenou \& Luri (this volume), to apply a parametric method
(the LM method in our case) it is necessary to model the physics and
observational errors of the sample as well as the censorship applied for
its construction. Each distribution (luminosity, velocity, spatial and
error distribution)  and the observational selection is described by a
p.d.f.  depending on a set of parameters.  The values of these parameters
are determined by applying the LM method to the sample. 

In our case we have used the modeling described in this section to
describe our sample of Hipparcos Cepheids. 

\subsection{Absolute magnitude distribution}

\begin{itemize}

  \item Linear mean PL relation:  $M_{v}=\alpha +\beta \; \log (P)$
  \item Gaussian dispersion ($\sigma _{M}$) of the individual
        absolute magnitude around the mean relation

        Actually, the real intrinsic distribution has a
        Top-Hat shape, but combined with the errors in
        the estimation of the interstellar absorption it
        is reasonably well approximated by a Gaussian.

  \item Interstellar absorption from the 3D model by Arenou et al. (1992)

\end{itemize}

\subsection{Velocity distribution}

\begin{itemize}
  \item Velocity ellipsoid: 
         \( \Phi _{K}(U,V,W)=e^{-\frac{1}{2}(\frac{U-U_{0}}{\sigma _{U}})^{2}}\; e^{-\frac{1}{2}(\frac{V-V_{0}}{\sigma _{V}})^{2}}\; e^{-\frac{1}{2}(\frac{W-W_{0}}{\sigma _{W}})^{2}} \)

  \item Galactic rotation: Oort-Lindblad model at first order with 
        \linebreak[4]
        $A=14.4 \; km \;s^{-1} \; kpc^{-1}$, 
        $B=-12.8 \; km \;s^{-1} \; kpc^{-1}$ and  $R_{\odot }=8.5 \; kpc$
        (Kerr \& Lynden-Bell, 1986).

\end{itemize}

\subsection{Spatial distribution}

\begin{itemize}
  \item Exponential disk: \( \Phi _{S}(x,y,z)=e^{-|\frac{z}{Z_{0}}|} \)
\end{itemize}

\subsection{Error distribution}

As we are dealing with an all-sky sample, we can safely assume that the
astrometrical observational errors of the stars are uncorrelated and
follow a Gaussian distribution (Arenou \& Luri, this volume). We have also
assumed Gaussian errors for radial velocity but we have considered
negligible the errors in apparent magnitude, position and period.

\subsection{Observational selection}

As we have included all the classical Cepheids contained in the Hipparcos 
Catalogue, the only observational selection is the one of the catalogue
itself.

\begin{figure}
\plotone{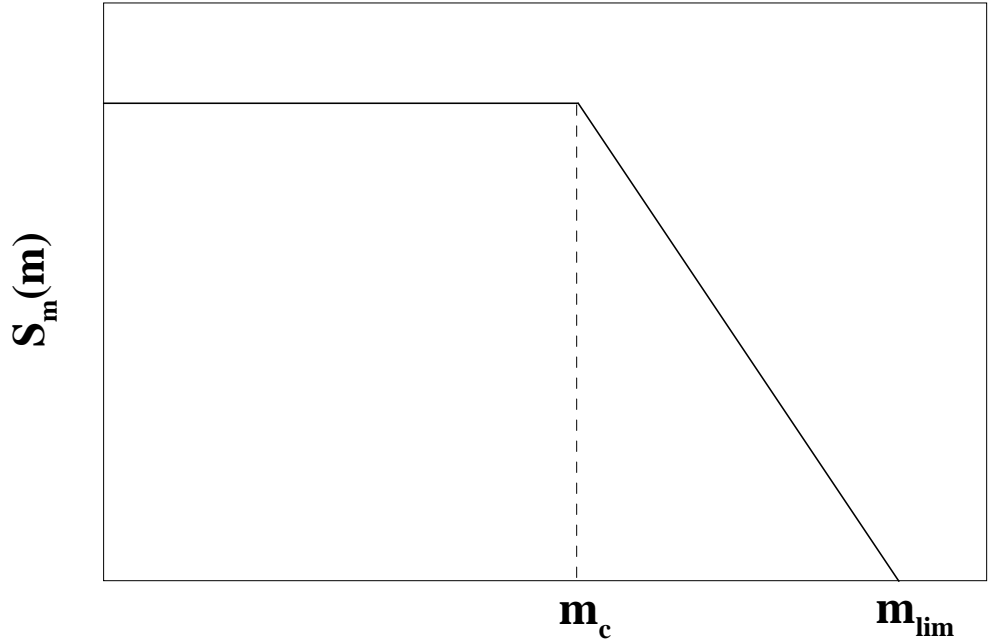}
\caption{Selection function in apparent magnitude} \label{fig_S}
\end{figure}

The main censorship of the catalogue is in apparent magnitude: the
catalogue is complete up to the Hipparcos magnitude $H_{p}\simeq 7.9^{m}$
(the Survey) and its magnitude limit is \( H_{p}\simeq 12.5^{m} \). We
have modeled this censorship with a selection function giving completeness
up to a certain apparent magnitude $m_c$ (estimated by the method togheter
with the rest of parameters) and a linear decrease of the completeness up
to the limiting magnitude (Figure \ref{fig_S}). 

%
%

On the other hand, we have considered the censorship on \( (\pi
_{t},l,b,\mu _{\alpha },\mu _{\delta },v_{r},P) \)  not significant on
average.

\subsection{Groups}

As pointed out in Sec. \ref{sec_sample}, our sample is a mixture of two
types of Cepheids: fundamental pulsators and overtone pulsators. We have
classified the stars in these two types using the Fourier coefficients
obtained from the Hipparcos light curves:  204 were classified as
fundamental pulsators and 34 as overtone pulsators. 

We have also made two supplementary hypothesis:

\begin{enumerate}

  \item The slopes of the PL relations for both groups have been 
        fixed to the values for the SMC obtained by the EROS team
        (Sasselov et al., 1997): $\beta _{F}=-2.72\pm 0.07$ (fundamental)
        and $\beta _{OV}=-3.46\pm 0.14$ (overtones). Only the
        zero points of the PL relations have been determined,
        $\alpha _{F}$ (fundamental) and $\alpha _{OV}$ (overtones).

  \item We have assumed that both groups have the same kinematic and
        spatial distributions (i.e. the same velocity ellipsoid and
        scale height).

\end{enumerate}

\section{Results}

The application to our sample of the LM method with the modeling described
above gives the following estimates for the zero points of the PL
relations:

\begin{description}

  \item[Cepheids pulsating on the fundamental mode:]

        \( \alpha _{F}=-1.01^m \pm 0.13 \)

  \item[Cepheids pulsating on the 1st overtone:]

        \( \alpha _{OV}=-1.26^m \pm 0.20 \)

\end{description}

\noindent On the other hand, the values obtained for the kinematical and
spatial distribution parameters are 
$(U_0=-9.0\pm1.5,V_0=-10.3\pm1.2,W_0=-7.5\pm0.6)$ $km\;s^{-1}$,
$(\sigma_U=13.0\pm1.2,\sigma_V=12.7\pm1.3,\sigma_W=6.2\pm1.2)$ $km\;s^{-1}$ 
and $Z_0=97\pm7$ $pc$.

\section{LMC distance modulus}

The LMC distance modulus can be calculated from the difference of zero
points between our PL relation and the EROS apparent magnitude PL relation
for the LMC (Sasselov et al. 1997). However, the difference of zero points
has to be corrected of the effects of interstellar absorption and
metallicity and thus two more assumptions are needed:

\begin{itemize}

  \item We have assumed a value of $E(B-V)=0.1^{m}$ for the LMC
        extinction (from Freedman et al., 1994). Then, we have 
        calculated the absorption from this value using the
        ratio $R=\frac{A_{v}}{E(B-V)}=3.311$

  \item We have assumed a correction of $0.043^{m}$ to take into
        account the effects on the zero points of the metallicity
        difference between the LMC and our Galaxy (Laney \& Strobie, 1994)

\end{itemize}

\noindent Using these assumptions the resulting distance modulus for
the LMC is:

\begin{center}
  \( (m-M)_0 = 18.35^{m}\pm 0.13 \)
\end{center}

This result is in agreement with the ``short" distance scale obtained from
(for instance) RR-Lyrae (Luri et al. 1998)  but not compatible with the
``long" distance scale from other Hipparcos results for Cepheids (Feast \&
Catchpole, 1997). Please notice that the error estimation does not take
into account the error in the value of $E(B-V)$ for the LMC.


\end{document}